\documentclass[sigconf,nonacm,screen,natbib=false,9pt]{acmart}

\AtBeginDocument{%
  \providecommand\BibTeX{{%
    Bib\TeX}}}

\AtBeginDocument{
    \hypersetup{
        colorlinks=true,
        citecolor=blue,      
        linkcolor=blue,    
        urlcolor=blue
    }
}
\RequirePackage[
  datamodel=acmdatamodel,
  style=acmnumeric,
  ]{biblatex}

\usepackage{pifont}
\usepackage{array}
\usepackage{graphicx}
\usepackage{subcaption}
\usepackage{multirow}
\usepackage{algorithm}
\usepackage{url}
\usepackage{enumitem}
\usepackage{pifont}

\providecommand{\wwnote}[1]{{\protect\color{red}\noindent {#1}}}

\begin{document}

\providecommand{\wwnote}[1]{{\protect\color{red}\noindent {#1}}}
\def\BibTeX{{\rm B\kern-.05em{\sc i\kern-.025em b}\kern-.08em
    T\kern-.1667em\lower.7ex\hbox{E}\kern-.125emX}}
\providecommand{\sznote}[1]{{\protect\color{blue}\noindent {#1}}}
\def\BibTeX{{\rm B\kern-.05em{\sc i\kern-.025em b}\kern-.08em
    T\kern-.1667em\lower.7ex\hbox{E}\kern-.125emX}}
    
\title{An End-to-End Distributed Quantum Circuit Simulator}
\author{Sen Zhang}
\affiliation{%
  \institution{George Mason University}
  \city{Fairfax}
  \state{VA}
  \country{USA}
}

\author{Lingjun Xiong}
\affiliation{%
  \institution{George Mason University}
  \city{Fairfax}
  \state{VA}
  \country{USA}
}

\author{Yipei Liu}
\affiliation{%
  \institution{George Mason University}
  \city{Fairfax}
  \state{VA}
  \country{USA}
}

\author{Brian L. Mark}
\affiliation{%
  \institution{George Mason University}
  \city{Fairfax}
  \state{VA}
  \country{USA}
}

\author{Lei Yang}
\affiliation{%
  \institution{George Mason University}
  \city{Fairfax}
  \state{VA}
  \country{USA}
}

\author{Zebo Yang}
\affiliation{%
  \institution{Florida Atlantic University}
  \city{Boca Raton}
  \state{FL}
  \country{USA}
}

\author{Weiwen Jiang}
\affiliation{%
  \institution{George Mason University}
  \city{Fairfax}
  \state{VA}
  \country{USA}
}








\begin{abstract}
\noindent Quantum computing has made substantial progress in recent years; 
however, its scalability remains constrained on a monolithic quantum processing unit (QPU). Distributed quantum computing (DQC) offers a pathway by coordinating multiple QPUs to execute large-scale circuits. Yet, DQC still faces practical barriers, as its realization depends on advances in hardware-level components such as quantum transducers and high-fidelity entanglement-distribution modules. While these technologies continue to improve, mature DQC platforms remain unavailable. In the meantime, researchers need to assess the benefits of DQC and evaluate emerging DQC designs, but the software ecosystem lacks a circuit-level simulator that models heterogeneous backends, noisy connections, and distributed execution. 
To fill this gap, this paper proposes \texttt{SimDisQ}, the first end-to-end circuit-level DQC simulator, composed of a set of novel DQC-oriented automated simulation toolkits and communication noise models that can interoperate with existing toolkits in mainstream quantum software ecosystems.
Leveraging circuit-level simulation capabilities, \texttt{SimDisQ} enables quantitative exploration of architectural design trade-offs, communication fidelity constraints, and new circuit optimization challenges introduced by DQC, providing a foundation for future research in this promising direction.
Benchmarking experiments using \texttt{SimDisQ} respond to a couple of open questions in the community; for example, noisy simulation of superconducting and trapped-ion qubits, with a reasonable entanglement-distribution fidelity, reveal that 
heterogeneous QPUs can indeed yield higher execution fidelity.








\end{abstract}

\maketitle

\section{Introduction}
\noindent Quantum computing has demonstrated remarkable progress across various domains, drawing increasing attention from both academia and industry \cite{aghaee2025scaling, cuomo2020towards,gambetta2020ibm, main2025distributed}. Leading tech companies, including IBM, Rigetti, IonQ, Quantinuum, and QuEra, have invested heavily in developing high-performance quantum processing units (QPUs) \cite{ibm2025dqc, ionq2025dqc}. However, the monolithic scaling of QPUs remains fundamentally constrained by fabrication challenges and the accumulation of physical errors \cite{materialschallenges}. This is analogous to the evolution of classical computing, where performance gains from single CPUs eventually plateaued, prompting a shift toward distributed architectures that ultimately established a key evolutionary pathway for modern computing systems \cite{corbett2013spanner, ghemawat2003google}. Likewise, distributed quantum computing (DQC) offers a scalable path forward by interconnecting multiple QPUs via optical switching networks \cite{main2025distributed}, enabling collaborative execution of large-scale quantum algorithms. Companies such as IBM, Cisco, and IonQ are actively pursuing interconnected, modular QPU architectures to construct larger quantum systems \cite{ibm2025dqc, ionq2025dqc, shapourian2025cisco}. This emerging paradigm extends the classical notion that \textit{``the network is the computer''} into the quantum era, where the computational fabric is no longer a single processor, but an entangled network of cooperating QPUs.

Despite the momentum, software support for DQC remains limited. Existing industry-grade software tools are designed almost exclusively for single-QPU execution. Transpiler frameworks such as Qiskit, Cirq, and Braket offer sophisticated optimization pipelines, but they cannot introduce remote gates or coordinate multi-QPU compilation. 
Noise simulators like Qiskit Aer, CUDA-Q, and Cirq model device-level noise but exclude inter-QPU communication, photon-loss channels, and topology-dependent errors. 
In academia, there have emerged quantum networking simulators, including SeQUeNCe \cite{wu2021sequence}, NetSquid \cite{coopmans2021netsquid}, and QuISP \cite{matsuo2019simulationdynamicrulesetbasedquantum}, that capture photonic links and entanglement distribution with high fidelity, yet remain disconnected from circuit-level compilation and circuit simulation. Even recent attempts at distributed execution \cite{diadamo2021distributed,ferrari2024dqcframework,pouryousef2025network}
implement only partial aspects of DQC, such as scheduling.
To the best of our knowledge, no existing framework can transform a monolithic circuit into a remote-gate-compatible, topology-aware distributed circuit while simultaneously accounting for device noise, communication noise, backend heterogeneity, and partitioning decisions.

In this paper, we present \texttt{SimDisQ}, an end-to-end DQC simulation and benchmarking framework designed to enable systematic exploration of distributed architectures. Moving beyond traditional monolithic transpilation and simulation, \texttt{SimDisQ} is built around four essential pillars required for DQC simulation: (1) a topology-aware constructor that organizes logical circuit structure according to multi-QPU network layouts; (2) a backend-specific isolator that enables independent transpilation tailored to native gates, coupling maps and noise characteristics of each QPU; (3) a deadlock-free assembler that integrates independently transpiled subcircuits by inserting and scheduling remote-gate operations; and (4) an integrated noise model that unifies QPU noise with optical-loss and distance-dependent communication noise. 

The main contributions of this paper are as follows:

\begin{itemize}[noitemsep,topsep=2pt]
    \item \texttt{SimDisQ} represents the first circuit-level simulator for distributed quantum computing, integrating seamlessly with the existing quantum software ecosystem.
    \item We perform a set of benchmarking experiments to demonstrate the necessity of DQC, evaluate DQC architecture design trade-offs, and reveal critical gaps in software stacks.
\end{itemize}

Using \texttt{SimDisQ}, we benchmark six representative circuits (e.g., quantum error correction code and variational quantum circuits) across five DQC architectures (e.g., heterogeneous QPUs in terms of size and qubit types).
Results reveal key insights: distributing circuits across multiple small QPUs can outperform a single, larger but noisier QPU in terms of fidelity under certain network topologies and backend configurations. Together, these findings underscore the need for dedicated circuit-level DQC simulation and demonstrate how \texttt{SimDisQ} facilitates the co-design of hardware architectures, network topologies, and compilation strategies for emerging DQC systems.

\begin{figure}[t]
  \centering
  \includegraphics[width=1\columnwidth]{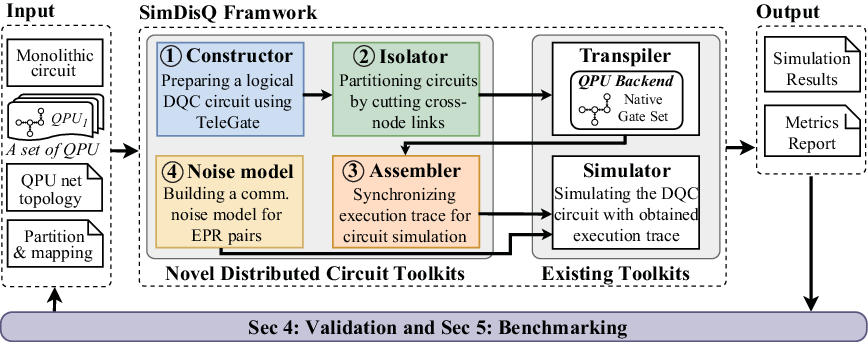}
  \caption{Overview of the \texttt{SimDisQ} Framework.}
  \label{fig:simdisq}
\end{figure}

\section{\texttt{SimDisQ} Framework}
Figure \ref{fig:simdisq} presents the proposed \texttt{SimDisQ} framework.  
As shown in this figure, \texttt{SimDisQ} contains four novel contributing toolkits that will interoperate with the existing toolkits, ensuring compatibility with today's quantum software ecosystems, including {\Large{\ding{172}}} constructor, {\Large{\ding{173}}} isolator, {\Large{\ding{174}}} assembler, and {\Large{\ding{175}}} noise model.
Specifically, the {\Large{\ding{172}}} constructor will take the user's monolithic circuit $C$ (i.e., traditional logical circuit for a single QPU), a set of QPUs $Q$, their connection topology $N$, and user-specified circuit partition $P$ and subcircuit-to-QPU mapping $M$ as input, which will generate a logical DQC circuit by replacing multi-qubit gates across QPUs with TeleGate.
Then, {\Large{\ding{173}}} isolator will cut the link on the cross-QPU gates (i.e., EPR pair and classical control gate) to prepare independent subcircuits, which can then be transpiled independently by existing tools.
Next, {\Large{\ding{174}}} assembler will take the transpiled subcircuits as input to generate an execution trace that will be used by the simulator.
Finally, {\Large{\ding{175}}} a noise model module for communication errors will be constructed, supporting noisy simulation.
The output of the \texttt{SimDisQ} framework includes the DQC circuit execution results and a detailed report on different performance and circuit metrics.

In the following, we will detail the four major components of \texttt{SimDisQ}.
Then, Sections \ref{sec:val} and \ref{sec:bench} will validate the simulation functionality and benchmark different circuits across distributed QPUs.


\noindent{\large\ding{192}} \textbf{Constructor}

\begin{figure}[t]
  \centering
  \includegraphics[width=1\columnwidth]{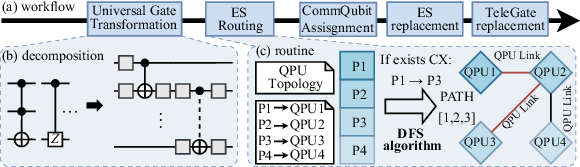}
  \caption{Illustration of the proposed five-step constructor.}
  \label{fig:Constructor}
\end{figure}

As shown in Figure \ref{fig:Constructor}(a), the proposed constructor includes five sequential steps.
We adopt the EPR pair for remote CNOT to be the communication primitive between subcircuits. 
In this step, we will first identify all remote gates $RG$ that are multi-qubit gates in $C$, where the corresponding qubits are assigned to different partitions $P$, and mapped to distributed QPUs in $Q$.
Then, for $\forall g\in RG$ and $g$ is not a CNOT gate, we will decompose $g$ into certain equivalent single-qubit gates and CNOTs \cite{nielsen2010quantum}, as shown in Figure \ref{fig:Constructor}(b), facilitating the use of TeleGate.
After the decomposition, we denote $CRG$ as the set of all remote CNOT gates.

Next, we apply ES according to the QPU topology $N$ and mapping $M$. Specifically, we will check $\forall g\in CRG$ if the control qubit and target qubit of $g$ are assigned to two QPUs ($Q_1$ and $Q_2$) without an optical channel.
If yes, a potential sequence of ES operations is needed to build EPR pairs between $Q_1$ and $Q_2$.
Since the EPR pair is noise-sensitive, the routing algorithm needs to select an optimal path with minimized resource consumption.
In this module, a depth-first search (DFS) algorithm is employed to identify the optimal path for ES, as shown for $QPU 1$ and $QPU 3$ in Figure \ref{fig:Constructor}(c). 

The next step is to allocate communication qubits in each QPU with an assigned partition (or subcircuit).
We apply a reset operation on these qubits after each EPR pair is consumed, allowing them to be reused and thus minimizing the number of communication qubits.
If a QPU needs to act as a repeater for ES, 
it requires two communication qubits; if a QPU is only involved to perform TeleGate for remote CNOT, 
it requires one communication qubit; otherwise, the QPU does not need any communication qubit.

In the last two steps of Figure \ref{fig:Constructor}(a), ES and TeleGate placements replace remote gates with ES and TeleGate circuits. 

\vspace{3pt}
\noindent{\large\ding{193}} \textbf{Isolator}

We develop an isolator module to cut cross-QPU links by using virtual gates $VG$ (i.e., instruction objects) and ensure the position of $VG$ will not be changed for optimization.

\begin{figure}[t]
  \centering
  \includegraphics[width=1\columnwidth]{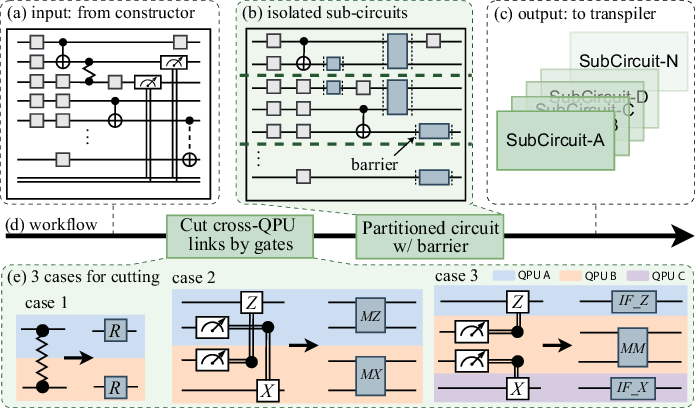}
  \caption{Illustration of the proposed sub-circuits isolator.}
  \label{fig:isolator}
\end{figure}

After ES and TeleGate replacement steps in the constructor, we obtain the logical DQC circuit. Using it as input (Figure \ref{fig:isolator}(a)), the isolator handles only three types of cross-QPU gates, as shown in Figure \ref{fig:isolator}(e).
To ensure deadlock-free assembly later, we cut the links by replacing these gates with virtual gates in three cases (see Figure \ref{fig:isolator}(e)). 
\textit{Case 1:} For EPR pair, we replace it by placing a $VG$ (e.g., $R$ gate) on each qubit.
\textit{Case 2:} For two pairs of measurement and conditional gates used in TeleGate, we will use a multi-qubit $VG$ (e.g., $MZ$, $MX$) on each QPU.
\textit{Case 3:} For two pairs of Measurement and conditional gates used in ES, we will use a multi-qubit $VG$ on the repeater QPU and a single-qubit $VG$ on the other two QPUs.


To prevent the transpiler from arbitrarily moving our inserted virtual gates during optimization, we insert \textit{barriers} before each custom instruction prior to transpilation, as shown in Figure \ref{fig:isolator}(b).

This will ensure that their positions remain fixed during optimization. 
After this, the isolated sub-circuits will be sent to the corresponding transpilers for transpilation, which is
the output of our isolator, as shown in Figure \ref{fig:isolator}(c).

\vspace{3pt}
\noindent{\large\ding{194}} \textbf{Assembler}

\begin{figure}[t]
  \centering
  \includegraphics[width=1\columnwidth]{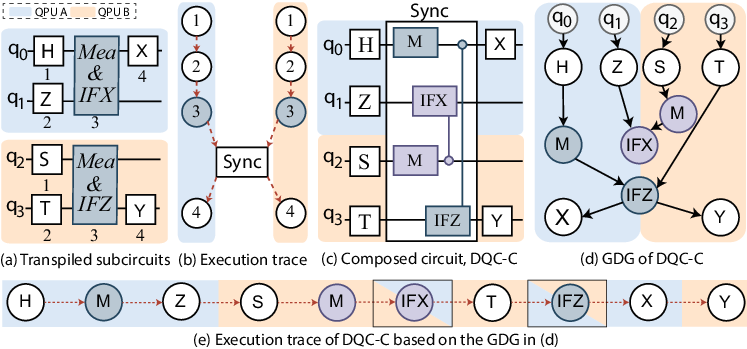}
  \caption{Workflow of assembler from transpiled subcircuits to a DQC circuit (DQC-C) with correct execution trace.}
  \label{fig:Assembler}
\end{figure}

Figure \ref{fig:Assembler} shows the workflow of the proposed assembler, 
which takes the transpiled subcircuits as input (in Figure \ref{fig:Assembler}(a)), and generates the composed circuit, DQC-C (see Figure \ref{fig:Assembler}(c)) with a correct execution trace (see Figure \ref{fig:Assembler}(e)) to enable the existing simulators for circuit-level simulation.

The assembler process is conducted in two steps.
The first step is to construct the circuit ``DQC-C''.
We will leverage the execution traces for isolated circuits.
Specifically, the assembler will follow the execution trace from each subcircuit to place the quantum gates in quantum registers from scratch.
If a node in the execution trace will not trigger a synchronization operation (say node 1 and 2 on QPU A in Figure \ref{fig:Assembler}(b)), we will place the associated gate on the corresponding qubit (i.e., $H$ on $q_0$ and $Z$ on $q_1$ in Figure \ref{fig:Assembler}(c)) .
Otherwise, we will block and wait at the synchronization point (say node 3 on QPU A or B) until all dependencies are satisfied.
In synchronization, there are three cases corresponding to the cutting of cross-QPU links in Figure \ref{fig:isolator}(e).
A reverse operation will be performed, which places the EPR pair, as well as the measurement and conditional gates, onto the corresponding qubits.
The sync block in Figure \ref{fig:Assembler}(c) shows the operation for Case 2.
By traversing all the nodes in the execution trace, we construct the final DQC circuit.

Note that the possible deadlock 
is prevented by the collaboration of isolator and assembler.
Specifically, for two measurement and conditional pairs,
instead of using four independent single-qubit VGs, our isolator employs two multi-qubit VGs (see Figure \ref{fig:Assembler}(a)). Then, the assembler processes two pairs at the synchronization point, which increases the assembler's complexity but can effectively avoid deadlocks. 

With the DQC circuit, the second step is to construct the gate dependency graph, as shown in Figure \ref{fig:isolator}(d), which can be implemented using the utility tools in the existing transpiler (e.g., circuit\_to\_dag in Qiskit).
After that, the topological order can be extracted by applying a breadth-first search (BFS) algorithm.
Figure \ref{fig:Assembler}(e) shows the execution order of the holistic DQC circuit.






\vspace{3pt}
\noindent{\large\ding{195}} \textbf{Communication-Integrated Noise Model}

To 
enable noisy simulation, there is a need to build a communication noise model for inter-QPU operations.
In the assembled circuit, there are two types of remote operations: (1) the EPR pair generation and (2) the transmission of classical measurement results for a conditional quantum gate.
Since there exist mature fault-tolerant classical networks, we assume there are no errors in the classical communication channel.
Then, we only need to build the noise model for EPR pair generation, which is implemented via photonic quantum channels.
To this end, we adopt the standard optical-loss model, which is widely used in quantum networking literature \cite{pant2019routing}.
Specifically, the channel transmissivity is defined as follows:
\begin{equation}
\small
    \eta \sim e^{-\alpha L}
    \label{eq:optical_loss}
\end{equation}
where $L$ is the channel length and $\alpha$ is the attenuation coefficient.

For realistic simulation of the assembled circuit $AC$, \texttt{SimDisQ} further considers the integration of computation noise models from heterogeneous backends, together with the communication noise model.
The integration includes the following two steps.
First, for each partition $p\in P$ that is mapped to QPU $q=M(p)$, we will retrieve the noise model of $q$ (either from the current calibration data of working QPUs or the historical calibration data from the retired QPUs).
Second, according to the transpilation results, we will assign the gate error to the corresponding qubits in circuit $AC$, and the communication error to EPR pairs between a pair of QPUs.

After this, the noise model will be sent to the simulator for a noisy simulation, which generates simulation results and a report with a set of metrics (e.g., the number of consumed EPR pairs).





\begin{figure*}[t] 
  \centering
  \includegraphics[width=1\textwidth]{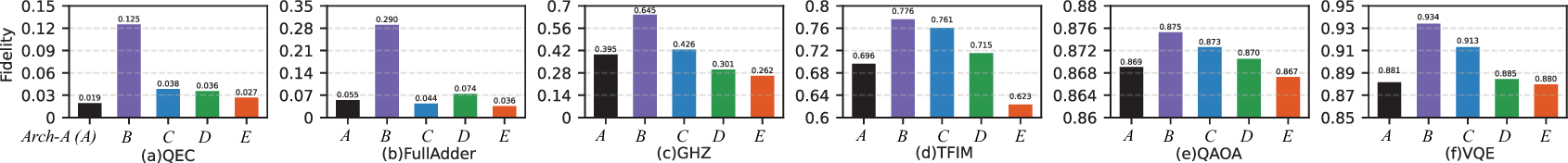} 
  \caption{Fidelity results across all benchmarks and architectures with an inter-QPU distance of 0.2km for data-center scale.}
  \label{fig:E_2}
\end{figure*}

\section{\texttt{SimDisQ} Framework Validation}\label{sec:val}


Before benchmarking, this section validates the correctness of the proposed \texttt{SimDisQ} under a noise-free environment, using a set of circuits that will also be used in the benchmarking in Section \ref{sec:bench}.
As shown in Table \ref{tab:Validation}, the target circuits cover representative applications in quantum computing, including (1) Steane code for quantum error correction (QEC), (2) a reversible full adder for arithmetic operations (FullAdder), (3) a GHZ-state generator for state preparation (GHZ); (4) a transverse-field Ising model for Hamiltonian simulation (TFIM); (5) Quantum Approximate Optimization Algorithm for optimization (QAOA); and (6) Variational Quantum Eigensolver for computing the ground state energy of a Hamiltonian (VQE).
Note that we implement the circuits of QEC and FullAdder following the standard design, while circuits of GHZ, TFIM, QAOA, and VQE are from the SupermarQ suite \cite{tomesh2022supermarq}.



The basic circuit properties, including qubits, depth, and 2-qubit gates, are presented in Table \ref{tab:Validation}.
To understand the complexity of partitioning these circuits across distributed QPUs, which is related to how densely the qubits interact with each other, we calcharacterize the circuit structure using interaction graph density (IGD), defined as $\rho = \frac{|E|}{n(n-1)/2}$, where $E$ is the set of unique qubit pairs connected by two-qubit gates and $n$ is the total number of qubits. 

For the purpose of validation, we implement the novel distributed circuit toolkits in Figure \ref{fig:simdisq} and the interface to access existing toolkits, i.e., IBM Qiskit, including its transpiler and simulator (Aer).
As \texttt{SimDisQ} will be validated on IonQ QPUs, we follow \cite{ionq_native_gates_qiskit} to ensure that the Qiskit transpiler supports IonQ's native gateset.

In Table \ref{tab:Validation}, the column of ``golden result'' gives the probability of a specific state (which has the largest amplitude) for each circuit.
``Single QPU'' reports the simulation of vanilla circuits without transpilation, while ``\texttt{SimDisQ}'' is our generated DQC circuits, both of which are simulated using Qiskit Aer simulator with 10k shots in a noise-free environment.
We can see that the deviations of simulation results from golden results are at the same level between Single QPU and \texttt{SimDisQ}.
This verifies the correct function of {\large\ding{192}} constructor, {\large\ding{193}} isolator, and {\large\ding{194}} assembler.


\begin{table}[t]
\centering
\scriptsize
\captionsetup{font=footnotesize}
\caption{Functionality validation of \texttt{SimDisQ}}
\label{tab:Validation}
\renewcommand{\arraystretch}{1.2}
\setlength{\tabcolsep}{2.5pt}
\begin{tabular}{|c|c|c|c|c|c|c|c|c|c|c|}
\hline
\multirow{2}{*}{\textbf{Circuit}}
& \multicolumn{4}{c|}{\textbf{Circuit Property}}
& \multicolumn{2}{c|}{\textbf{Golden result}}
& \multicolumn{2}{c|}{\textbf{Single QPU}}
& \multicolumn{2}{c|}{\textbf{\texttt{SimDisQ}}}\\
\cline{2-11}
& qubits & depth & 2Q & IGD & state & prob. 
& prob. & deviation & prob. & deviation \\
\hline
QEC & 13 & 15 & 35 & 0.45 & $|57\rangle$ & 0.125 & 0.131 & $10^{-4}$  & 0.1281 & $10^{-4}$  \\
FullAdder & 12 & 124 & 81 & 0.24 & $|910\rangle$ & 1 & 1 & 0  & 1 & 0   \\
GHZ & 16 & 17 & 15 & 0.13 & $|0\rangle$ & 0.5 
& 0.5046 & $10^{-5}$  & 0.4977 & $10^{-5}$   \\
TFIM & 12 & 55 & 66 & 0.17  & $|4095\rangle$ & 0.280 & 0.2837 & $10^{-2}$  & 0.2811 & $10^{-2}$   \\
QAOA & 8 & 54 & 56 & 1.00  & $|60\rangle$ & 0.020 & 0.0203 & $10^{-2}$  & 0.0197 & $10^{-2}$   \\
VQE & 10 & 14 & 9 & 0.2  & $|6192\rangle$ & 0.0009
& 0.0011 & $10^{-2}$  & 0.0011 & $10^{-2}$   \\
\hline
\end{tabular}
\end{table}

\begin{figure}[t]
  \centering
  \includegraphics[width=1\columnwidth]{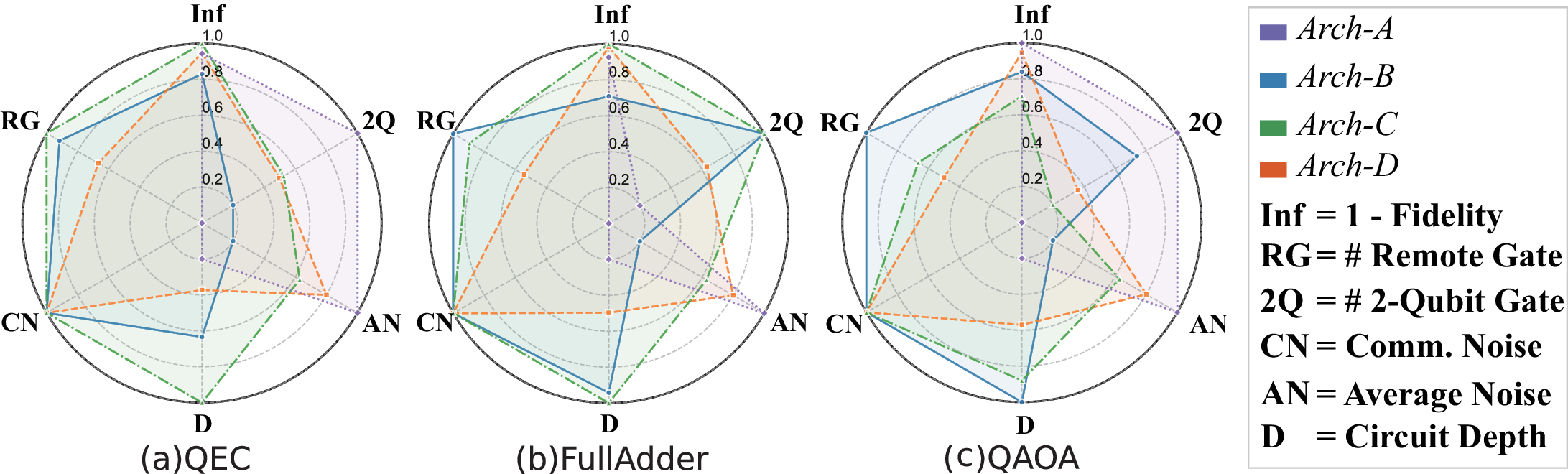} 
  \caption{Multi-metric evaluation of four DQC architectures
  }
  \label{fig:E_1}
\end{figure}

\vspace{-5pt}
\section{Benchmarking}\label{sec:bench}

To apply \texttt{SimDisQ} for benchmarking, we consider five architectures (A–E) that capture representative design trade-offs in DQC (see Table \ref{tab:BackendConfig}). For \textit{Arch-B} through \textit{Arch-E}, the hyperparameter $\alpha$ in Eq.~\ref{eq:optical_loss} is set to 0.05, following standard quantum-network modeling practices in \cite{azuma2023quantum}. The inter-QPU distance is configured at data-center scale (0.2 km), corresponding to typical intra-data-center link lengths.






We first carry out a holistic comparison of circuit-level metrics and fidelity across the the single-QPU architecture (\textit{Arch-A}) and three DQC architectures (\textit{Arch-B} to \textit{Arch-D}) with an inter-QPU distance of 2km, using three benchmarks with the highest IGD scores (i.e., QEC, FullAdder, and QAOA).
Figure \ref{fig:E_1} presents the result, where the north point reports infidelity so that smaller areas in the radar plots indicate better metrics. 
The results provide several key insights for DQC.
First, DQC architectures can outperform single-QPU in terms of fidelity: From all benchmarks, \textit{Arch-B} has lower infidelity over \textit{Arch-A}, and \textit{Arch-A} has the highest infidelity on QAOA in Figure \ref{fig:E_1}(c).
Second, the average gate noise (AN) has a strong correlation to fidelity: Even \textit{Arch-B} generally has a higher value for other metrics, but it has the lowest AN, leading to lower infidelity.
Third, heterogeneous DQC can also bring benefits. For QAOA, \textit{Arch-C} with IBM and IonQ QPUs outperforms all other architectures, as it provides a balanced tradeoff across all metrics (more insights are provided in Section 5.3).

To better compare fidelity across all circuits in Table \ref{tab:Validation}, we carry out a set of benchmarks for DQC built in a data-center scale, where the distance is set to 0.2km.
With such a setting, the results in Figure \ref{fig:E_2} show that \textit{Arch-B} achieves the best performance on all benchmarking circuits, revealing that leveraging multiple high-fidelity (albeit small) devices can outperform a larger but noisier backend, challenging the common assumption that minimizing circuit partitioning inherently yields better performance.

The results also reveal several noteworthy observations.  First, by bringing a higher fidelity trapped-ion qubit system, the heterogeneous architecture \textit{Arch-C} achieves the second-best fidelity, closely following \textit{Arch-B} on all benchmarks except FullAdder.
Second, by pairing the high-noise QPU (i.e., \textit{Arch-A}) with a smaller but high-fidelity QPU, \textit{Arch-D} always performs better than \textit{Arch-A}.
These observations quantitatively demonstrate the effectiveness of heterogeneous DQC.
Together with the superior results of \textit{Arch-B}, benchmarking results clearly exhibit the necessity of DQC.

\begin{table}[t]
\centering
\scriptsize
\captionsetup{font=footnotesize}
\caption{Architectures used for benchmarking}
\label{tab:BackendConfig}
\renewcommand{\arraystretch}{1.08}
\setlength{\tabcolsep}{0.92pt}
\begin{tabular}{|c|c|c|c|>{\centering\arraybackslash}m{3.5cm}|}
\hline
  \textbf{Arch.} & \textbf{\# QPUs} & \textbf{\# qubits} & \textbf{Backend Types} & \textbf{Description} \\
\hline
A & 1 & 28 & ibm\_cambridge & Monolithic QPU (larger but noisier) \textbf{This is benchmarking baseline} \\
\hline
B & 4 & 5+5+5+5 & ibm\_vigo $\times$ 4 & Homogeneous fully-connect QPUs (smaller, less noisy, more networking) \\
\hline
C & 2 & 28 + 25 & ibm\_cambridge + IonQ\_Aria 2 & Heterogeneous in terms of QPU types\\
\hline
D & 2 & 28 + 5 & ibm\_cambridge + ibm\_vigo & Heterogeneous in terms of \# qubits \\
\hline
E & 2 & 28 + 28 & ibm\_cambridge $\times$ 2 & Homogeneous large-QPU backends \\
\hline
\end{tabular}
\end{table}

\vspace{5pt}
\section{Conclusion}

As QPUs struggle to scale up for qubit-intensive applications, DQC has become crucial for overcoming the limitations of monolithic devices. 
This paper presents \texttt{SimDisQ}, the first circuit-level DQC simulator that integrates architectural modeling, heterogeneous transpilation, and inter-QPU noise modeling. 
Our benchmarking results address several key issues in the community, specifically regarding DQC advantages, architectural trade-offs, and gaps in DQC software. These findings establish \texttt{SimDisQ} as a practical framework for future DQC research and design.




\printbibliography










\end{document}